\documentclass[
 reprint,
 aps,
 superscriptaddress
]{revtex4-1}

\usepackage{graphicx}

\begin{document}

\title{
Tuning the size and configuration of nanocarbon microcapsules:\\aqueous method using optical tweezers}
\author{Hiroshi Frusawa}
\email{frusawa.hiroshi@kochi-tech.ac.jp}
\author{Youei Matsumoto}
\affiliation{Institute for Nanotechnology, Kochi University of Technology, Tosa-Yamada, Kochi 782-8502, Japan.}


\begin{abstract}
To date, optical manipulation techniques for aqueous dispersions have been developed that deposit and/or transport nanoparticles not only for fundamental studies of colloidal dynamics, but also for either creating photonic devices or allowing accurate control of liquids on micron scales.
Here, we report that optical tweezers (OT) system is able to direct three-dimensional assembly of graphene, graphite, and carbon nanotubes (CNT) into microcapsules of hollow spheres.
The OT technique facilitates both to visualize the elasticity of a CNT microcapsule and to arrange a triplet of identical graphene microcapsules in aqueous media.
Furthermore, the similarity of swelling courses has been found over a range of experimental parameters such as nanocarbon species, the power of the incident light, and the suspension density.
Thanks to the universality in evolutions of rescaled capsule size, we can precisely control the size of various nanocarbon microcapsules by adjusting the duration time of laser emission.
\end{abstract}

\maketitle

Carbon atoms arranged in a honeycomb-like lattice constitute the common building blocks of not only graphite but also other carbon nanomaterials including fullerene, carbon nanotubes (CNTs), and graphene \cite{assembly-book, graphene}.
The individual particles of these nanocarbons possess unique mechanical, electronic, and optical properties, and there is a variety of assembly techniques for displaying these properties macroscopically or mesoscopically \cite{assembly-review1,capsule-review, hybrid}.
Correspondingly, the hollow-caged superstructure of the graphene or CNT assembly has been fabricated in several ways \cite{capsule-review,capsule-graphene1,capausle-graphene2,capsule-graphene3,capsule-shinkai,capsule-emulsion1,capsule-shi,opt-capsule, capsule-lumi, capsule-IR, capsule-sensing}.
The majority of methods developed for these capsular organizations are classified as layer-by-layer (LbL) assembly techniques \cite{capsule-review, capsule-graphene1,capausle-graphene2,capsule-graphene3,capsule-shinkai,capsule-emulsion1,capsule-shi,capsule-IR,capsule-sensing}.
The LbL method includes the step of depositing charged nanocarbon particles on spherical templates, such as polymer colloids and water droplets in oil, and necessarily involves a final step to remove the template cores \cite{capsule-review, capsule-graphene1,capausle-graphene2,capsule-graphene3,capsule-shinkai,capsule-emulsion1,capsule-shi,capsule-IR,capsule-sensing}.
There is also a recent report that CNT assembly is optically directed to form spheres on the ends of optical fibers without the addition of core materials \cite{opt-capsule}.

\begin{figure*}[ht]
        \begin{center}
	\includegraphics[
	width=12 cm
	]
       {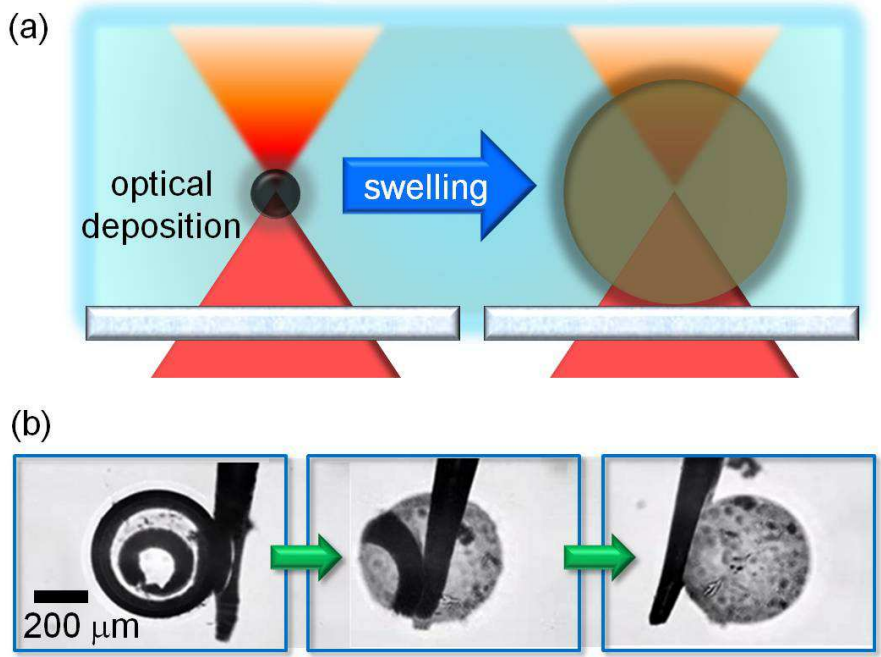}
	\end{center}
	\caption{{\bf A schematic of optical swelling and a CNT microcapsule manipulated}.
(a) Diagrammatic comparison between two typical situations of nanocarbon dispersions in optical tweezers system illustrates the collection of nanocarbon dispersoids at the focal point, located above the glass substrate, and the isotropic swelling while continuing to add nanocarbons on the outer surface.
(b) Optical images of a CNT microcapsule adhering to the microneedle. The isotropic structure of a spherical shape is determined from observing the various sides of the rotating sphere.}
\end{figure*}

Our concern relates to the last finding of optically directed three-dimensional (3D) assembly \cite{opt-capsule}, which is further traced back to the optical deposition of CNTs or graphene on the end of an optical fiber \cite{deposition-1, deposition-monitor,deposition-2,graphene-deposition, graphene-device1,graphene-device4, graphene-deposition2}.
It has been demonstrated that these aggregates themselves can form photonic devices with fascinating properties such as an ultrafast recovery time, ultrafast saturable absorbance, and optical nonlinearity \cite{deposition-1, deposition-monitor,deposition-2,graphene-deposition, graphene-device1,graphene-device4, graphene-deposition2, graphene-cnt-device};
therefore, capsular structures may have potential applications in electrochemical or photonic devices, as well as in drug delivery and biosensors \cite{capsule-lumi, capsule-IR,capsule-sensing}.
Although the simplicity of the optical process is expected to facilitate practical applications, there remain both technical and fundamental issues as follows:
(i) The optical fiber system used previously \cite{opt-capsule, deposition-1, deposition-monitor,deposition-2,graphene-deposition, graphene-device1,graphene-device4, graphene-deposition2} has difficulty in both fabrication and manipulation, such that the CNT sphere adheres to the end of the fiber during the 3D assembly and must be removed to repeat the optical fabrication using the same fiber \cite{opt-capsule}.
The existing optical techniques \cite{opt-capsule,deposition-1, deposition-monitor,deposition-2,graphene-deposition, graphene-device1,graphene-device4, graphene-deposition2} are thus unsuitable for both the sequential processing and the position adjustment.
(ii) Furthermore, the growth processes need to be clarified so that we may manage the optical swelling; it seems difficult to precisely control the capsule size, if it is the microbubble generation by local heating \cite{opt-capsule, gold, metallic, gold2, vapor-1,vapor-plasma, vapor-2, vapor-3} that plays an essential role in forming the 3D assembly.

Here we settle these problems together using optical tweezers (OT) \cite{ot-grier, ot-review, ot-sort, ot-book, trap-grier, trap-1,trap-2,trap-3, trap-4, gtrap1, gtrap2}.
Optical tweezing is not only among the most established and popular methods for optical trapping and manipulation of micro-and nano-objects including CNTs and graphene flakes \cite{trap-grier, trap-1,trap-2,trap-3, trap-4, gtrap1, gtrap2}, but also provides emerging applications for analysis or transportation including weak force measurements, particle sorting, and optofluidics combined with the microfluidic toolkit \cite{ot-review, ot-sort, ot-book}.
In contrast to the fiber system, the OT technique is free of any restrictions on the controllability of the distance between the focal point and the surrounding objects, such as the cover slip and a microneedle inserted for manipulations.
Taking full advantage of the versatility of this technique, the technical limitations associated with the fiber systems could be overcome. 

In what follows, we aim to demonstrate that the OT method is capable of tuning the size and configuration of nanocarbon capsules;
for instance, a desired configuration of microcapsule arrays is created in aqueous media.
Furthermore, the controllability of capsule size is validated from investigating a variety of swelling kinetics in dispersed nanocarbons such as CNTs, graphenes and graphites.
In addition to the nanocarbon species, we have changed the experimental parameters including the solvent temperature $T_s$, the incident laser power $I$, and the density $\rho$ of nanocarbon dispersoids.

\subsection*{Results}
{\bf Manipulating nanocarbon microcapsules}. 
The technical point, which we have discovered, is to place the laser focus of OT system above an ordinary cover slip on which a drop of nanocarbon suspension is mounted, so that one could find that a spherical nanocarbon assembly, whose center is located at the focal point, is swelling larger as the duration time of laser emission is longer (see Fig.1(a) as a  schematic of the growth process).
We have adhered a swollen CNT sphere to a tungsten microneedle by the position adjustment of both the microcapsule and microneedle using the OT system and micromanipulator, respectively, which enables one not only to lift the attached object from the substrate by positional control of the microneedle, but also to rotate the CNT sphere together with the attached microneedle (see Fig. 1(b) and Supplementary Movie 1).
Thanks to the in situ observations, the 3D structure of spherical shape has been clarified.

\begin{figure}[ht]
        \begin{center}
	\includegraphics[
	width=8.8 cm
	]
       {{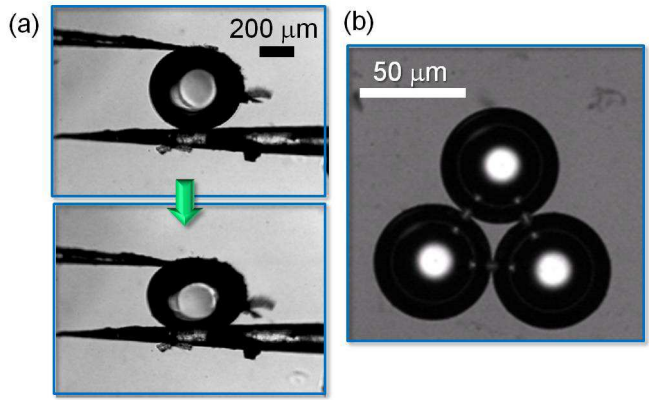}}
	\end{center}
	\caption{{\bf CNT and graphene microcapsules deformed and arranged}.
(a) A sequence of micrographs of a sandwiched CNT microcapsule before and after the deformation caused by narrowing the interspace between the microneedle pair. (b) An optical image of graphene microcapsule arrays. The triplet is arranged by successively focusing on three different points at regular intervals.}
\end{figure}
\begin{figure*}[ht]
     \begin{center}
	\includegraphics[
                width=13.5 cm
	]
{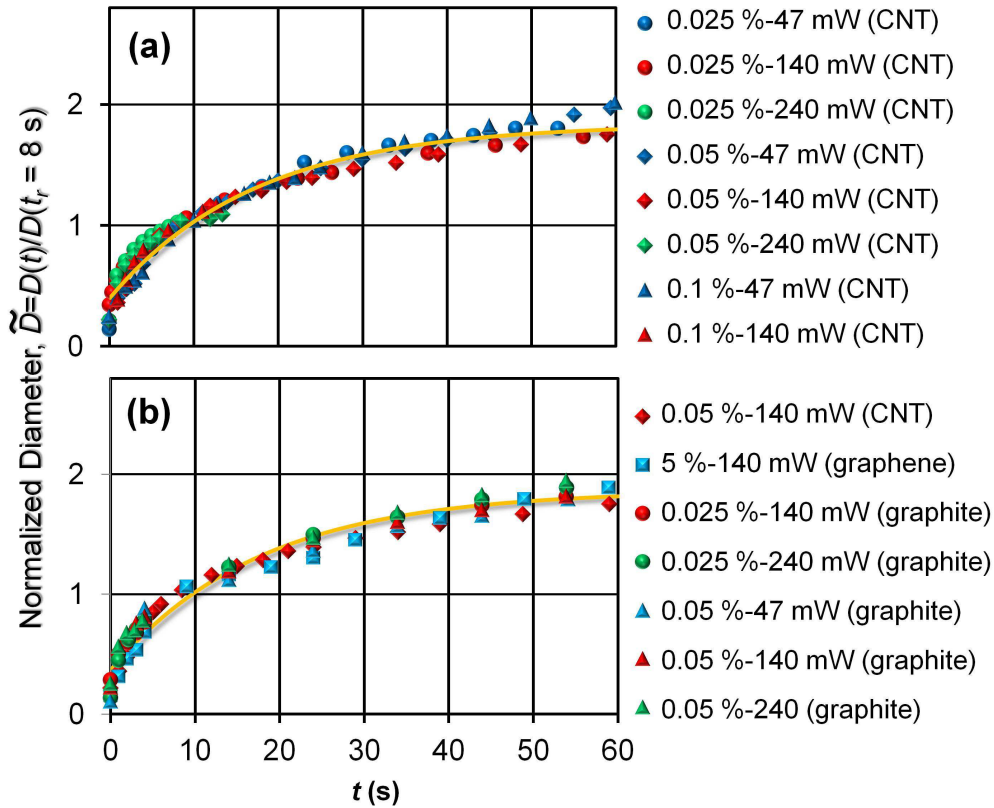}
	\end{center}
	\caption{{\bf A generic kinetics of various nanocarbon microcapsules}.
(a, b) Rescaled optical swelling behaviors are collected using the normalized diameter $\widetilde{D}(t)$ as a function of time $t$. We show the various growth processes in CNT suspensions of different densities for changes in the power of the incident light (a), and for distinct nanocarbon materials (b). The number pairs connected by hyphens represent the weight densities (left numbers) of dispersions and the incident light powers (right numbers) at the focal point, respectively. The fitting results are delineated by the curved lines.}
\end{figure*}

In Fig. 2(a) and Supplementary Movie 2, we see that capsule deformation is induced by exerting mechanical stress on a sandwiched CNT microcapsule via a pair of microneedles.
The relaxation to the initial isotropic state, observed in Movie 2, exhibits that the elasticity is imparted by the void space.
It has also been found from long-term observation that confined 3-$\mu$m colloids continue to undergo Brownian motion without permeating the carbon shell (see Supplementary Movie 3), which suggests no holes of several microns on the spherical surface.

The OT system is also able to create a triangular array of graphene microcapsules (Fig. 2(b)) which are detachable from each other, by adjusting the focal position independently of that of the cover slip.
These microcapsules have an almost equal diameter, because they are fabricated by changing the focal position from one point to another at regular intervals with the laser path shuttered while maintaining the solvent temperature $T_s$ at $T_s=25\,^\circ\mathrm{C}$.

{\bf A universal behavior of growth processes}. 
For systematically accomplishing the size-control, various growth processes need to be quantified;
therefore, we have measured the diameter $D(t)$ of a swelling microcapsule at a time specified by $t$, which is almost equal to the duration of the laser emission.
Strictly speaking, the initial time, $t=0$, was operationally defined by the instant when the CCD camera used for the long-term observations began to detect an optically deposited nucleate.

As a minimal model to reproduce the swelling kinetics of nanocarbon microcapsules, we consider a typical equation of relaxational phenomena: 
   \begin{equation}
   \frac{dD(t)}{dt}=-\frac{1}{\tau}D(t)+v_{\mathrm{dc}}
   \label{minimal},
   \end{equation}
whose solution is given by
   \begin{eqnarray}
   D(t)&=&D_{\mathrm{max}}\left[
   1-\exp\left\{-(t+t_0)/\tau\right\}
   \right],
   \label{solution}\\
   D_{\mathrm{max}}&=&
   \tau v_{\mathrm{dc}}.
   \label{max}
   \end{eqnarray}
Here $\tau$ and $v_{\mathrm{dc}}$ correspond to the relaxation time and the DC component of the swelling velocity, respectively, $D_{\mathrm{max}}$ denotes the maximum diameter at saturation, and $t=-t_0$ the emergence time defined mathematically: $D(-t_0)=0$.
Equation (1) describes the competition between the DC component $v_{\mathrm{dc}}$ of driving velocity for swelling and the declining rate $1/\tau$ against expansion.
Correspondingly, the maximum diameter $D_{\mathrm{max}}$ is given by the ratio of $v_{\mathrm{dc}}$ to $1/\tau$ (see eq. (\ref{max})).

As found from eq. (\ref{solution}), the time evolution of rescaled size $\widetilde{D}(t)$ defined by $\widetilde{D}(t)\equiv D(t)/D(t_r)$ depends only on time parameters of $\tau$, $t_0/\tau$, and an arbitrary reference time $t_r$:
   \begin{equation}
   \widetilde{D}(t)=\frac{1-\exp\left\{-(t+t_0)/\tau\right\}}{1-\exp\left\{-(t_r+t_0)/\tau\right\}},
   \label{rescaling}
   \end{equation}
which implies that a universal curve of $\widetilde{D}(t)$ is delineated by collecting various growth processes with similar times of both emergence and relaxation.

The dynamic universality, predicted by eq. (\ref{rescaling}), has been corroborated by gathering various swelling processes at $T_s=25\,^\circ\mathrm{C}$ in Fig. 3 where the normalized size $\widetilde{D}(t)$ is plotted consistently using an identical reference time $t_r$.
We set that $t_r=8$ sec. because growing CNT microcapsules at the maximal power of 240 mW became unstable beyond the present duration time and showed a sudden drop in diameter prior to saturation (see Fig. 4 and Supplementary Movie 4).

It is remarkable that Fig. 3 exhibits the generic kinetics at $T_s=25\,^\circ\mathrm{C}$, which is common to the incident laser power, suspension density and nanocarbon species.
We can describe the universal behavior by applying eq. (\ref{rescaling}) to the rescaled length as before, and both fitting curves of Figs. 3(a) and 3(b) obtain $\tau=17$ sec. and $t_0/\tau=0.2$.
In Supplementary Tables S1 and S2, we have fitting results for a variety of $D(t)$ without rescaling, which are in the range of $15\leq\tau\leq21$ and $0.2\leq t_0/\tau\leq0.3$ when incident laser powers, suspension densities and nanocarbon species are varied.
These variances of $\tau$ and $t_0/\tau$ underlie the universality depicted in Fig. 3.
In other words, the rescaled plot of $\widetilde{D}(t)$ facilitates to discriminate the variance in $t_0$ and/or $\tau$.

The duration of the laser emission at $D(0)$ was also determined, using a fast CMOS camera, which was approximately two seconds consistently with the fitting result of $t_0$;
incidentally, the fast CMOS camera detected the emergence of carbon nucleates after a mean laser duration of about one second irrespectively of experimental conditions.

{\bf Swelling mechanism}.
Both the symmetric shape and the hollow space of more than 100 $\mu$m in diameter reveal that the OT system is exerting an isotropic pressure over a long-ranged scale around the focal point, which suggests the existence of an optofluidic flow other than the inherent anisotropic force due to the optical gradient or scattering \cite{ot-sort}.
We investigate the details of $D(t)$ without use of the rescaled plot so that it may be explored what are determining factors of inducing the optofluidic flow.

\begin{figure}[hbtp]
        \begin{center}
	\includegraphics[
	width=8.5 cm
	]
{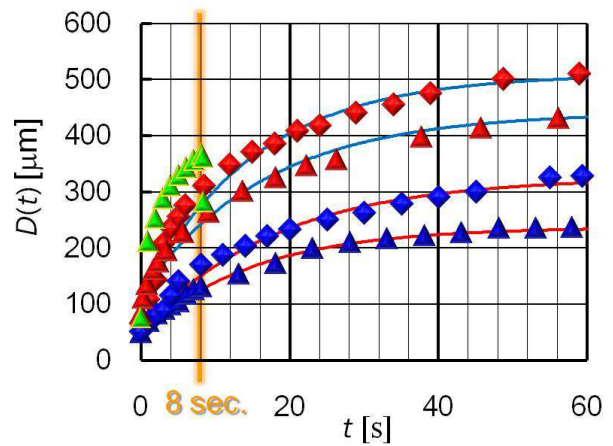}
	\end{center}
\caption{{\bf Time dependencies of capsule diameter on incident light power and suspension density}.
Measured diameter $D(t)$ as a function of $t$ in a CNT suspension.
The growth depends on the experimental parameters of both incident light power $I$ and suspension density $\rho$, where different data sets denote the following: blue triangles ($I_1,\rho_L$), blue diamonds ($I_1,\rho_H$), red triangles ($I_2,\rho_L$), red diamonds ($I_2,\rho_H$), green triangles ($I_3,\rho_L$), where $I_1=47$ mW, $I_2=140$ mW, $I_3=240$ mW, $\rho_L=0.025$ wt$\%$, and $\rho_H=0.05$ wt$\%$.
The curved lines are delineated for showing the fitting results.
At the maximum power of $I_3=240$ mW, the halfway capsule becomes unstable or breaks at around $t=8$ sec. as marked (see also Supplementary Movie 4).
}
\end{figure}

Figure 4 shows the change of $D(t)$ when the incident laser power $I$ at the focal point or the suspension density $\rho$ varies, from which we extrapolate that the capsule diameter at saturation is larger with increase of $I$ and/or $\rho$.
Supplementary Tables S1 and S2, the fitting results using eq. (2), actually verify that $D_{\mathrm{max}}$ is increased by raising $I$ and is slightly larger for a higher density of CNTs, whereas $v_{\mathrm{dc}}$ is enhanced by increasing $I$ but is weakly dependent on $\rho$;
at the maximal power of $I=240$ mW, we have found it difficult to appropriately fit eq. (2) to the corresponding data, due to the capsule breaking at an intermediate stage (see Supplementary Movie 4 as well as Fig. 4).

The $v_{\mathrm{dc}}$-dependencies on $I$ and $\rho$ can be explained by the change of temperature difference $T_f-T_s$ between the focal temperature $T_f$ and the solvent temperature maintained at $T_s=25\,^\circ\mathrm{C}$.
It has been found that the amount of absorbed light increases with increase of $I$ before reaching the saturation limit \cite{cnt-limiting, graphene-limiting}, as well as nanocarbon density of the focal deposition which is not necessarily proportional to the bulk concentration $\rho$.
It follows that the focal temperature $T_f$ becomes higher while the solvent temperature remains $T_s=25\,^\circ\mathrm{C}$, thereby creating a larger thermal gradient due to the increase of $T_f-T_s$.
Accordingly, the aforementioned increment of $v_{\mathrm{dc}}$ is ascribable to the steepening of thermal gradient \cite{gold, metallic, gold2, thermo-piazza, thermo-gold, thermo-review} which is caused by elevating $I$ and is induced to some extent by increasing $\rho$;
it is to be noted that the marginal $\rho$-contribution arises not only from the nonlinear relation between the bulk and deposited densities, but also from the enhancement of depositing flow with increase of $\rho$, which is in the opposite direction of $v_{\mathrm{dc}}$.

We also changed the solvent temperature $T_s$, as well as $T_f$.
Figure 5 compares the time dependencies of $D(t)$ in CNT suspensions at $T_s=25\,^\circ\mathrm{C}$ and $T_s=60\,^\circ\mathrm{C}$ with the other parameters equally set to $I=96$ mW and $\rho=0.05\,\mathrm{wt}\%$.
The optical micrograph in Fig. 5 represents an initial state at $T_s=60\,^\circ\mathrm{C}$ where the duration of the laser emission, determined using a fast CMOS camera, was approximately two seconds as mentioned before, which displays that the capsule interior is filled with microbubbles in the initial stage.
This microscopy image in Fig. 5 also exhibits not only a minimal size of generated microcapsules, but also a shell thickness of approximately 20 $\mu$m.
The apparent generation of microbubbles has revealed that the CNT dispersoids actually absorb light and raise the focal temperature $T_f$ in the present experimental conditions, as reported so far \cite{opt-capsule, vapor-plasma, vapor-2, cnt-limiting}.
In contrast, the optical microscopy of CNT microcapsules at $T_s=25\,^\circ\mathrm{C}$ is unable to distinguish microbubbles even at the maximal laser power (see Supplementary Movie 4).
Despite the clear difference in observations, these growth processes have a common feature that the increasing rate of $D(t)$ is explosively large at the initial stage, and subsequently slows down toward the saturated state.

\begin{figure}[hbtp]
        \begin{center}
	\includegraphics[
	width=8.8 cm
	]
{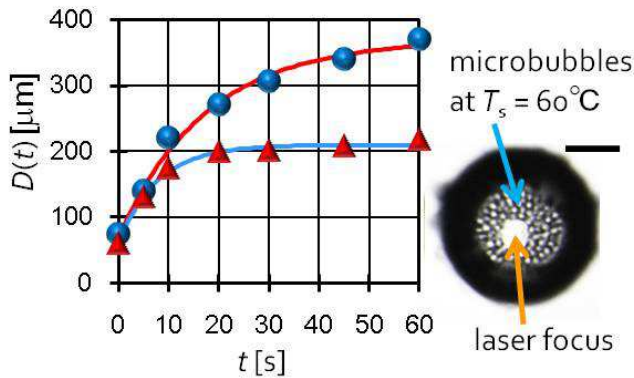}
	\end{center}
	\caption{{\bf Solvent temperature effect on the swelling kinetics}.
Measured diameter $D(t)$ as a function of time $t$ in a CNT suspension with a concentration of 0.05 wt$\%$ at different solvent temperatures of $T_s=25\,^\circ\mathrm{C}$ (blue circles) and $T_s=60\,^\circ\mathrm{C}$ (red triangles). The curved lines correspond to the fitting results. The optical micrograph shows an initial state of a CNT microcapsule at $T_s=60\,^\circ\mathrm{C}$ where laser focus and microbubbles are observed as indicated. Scale bar: 20 $\mu$m.}
\end{figure}

Despite the clear difference in observations, these growth processes have a common feature that the increasing rate of $D(t)$ is explosively large at the initial stage, and subsequently slows down toward the saturated state.
Comparing the long-time behaviors of $D(t)$ at different solvent temperatures in Fig. 5, one finds that $D_{\mathrm{max}}$ is smaller in a higher-temperature solvent ($T_s=60\,^\circ\mathrm{C}$).
According to the fitting results in Supplementary Table S3, $v_{\mathrm{dc}}$ is scarcely varied, whereas $\tau$ is reduced larger in correspondence with the decrease of solvent viscosity $\eta$ due to the increase of $T_s$:
we have $\tau_{25}/\tau_{60}\approx\eta_{25}/\eta_{60}\approx 2$, with the subscripts denoting the solvent temperatures.
This means that the reduction of the maximum diameter $D_{\mathrm{max}}$ is mainly caused by the viscosity diminishment associated with the $T_s$-change.
We thus find that the increase of $T_s$ makes a little change in the temperature difference $T_f-T_s$, and that the existence of microbubbles is not essential in enhancing the capsule growth beyond the viscosity reduction due to the increase of $T_s$, which is consistent with the previous observations of microbubbles as scavengers of dispersed CNTs \cite{opt-capsule, vapor-plasma, vapor-2}.

\subsection*{Discussion}

To summarize, the OT system has been found to be a fascinating tool for tuning the size and configuration of nanocarbon microcapsules in aqueous media, which is ensured by a universal behavior of swelling courses that applies to a range of incident light power $I$, nanocarbon density $\rho$, and nanocarbon species as displayed by Fig. 3.
Our finding of the generic kinetics enables one to efficiently control the capsule size by interpolating the appropriate duration time of laser emission even in an intermediate condition of $I$ and/or $\rho$.

For addressing the swelling mechanism in more detail, one should note the fitting results in Fig. 5 (see also Supplementary Table S3).
This indicates that the reduction of the maximum diameter $D_{\mathrm{max}}$ is mainly caused by the viscosity diminishment associated with the $T_s$-change.
In other words, the increase of $T_s$ makes a little change in the temperature difference $T_f-T_s$, and that the existence of microbubbles is not essential in enhancing the capsule growth beyond the viscosity reduction due to the increase of $T_s$, which is consistent with the previous observations of microbubbles as scavengers of CNTs, instead of the dispersers \cite{opt-capsule, vapor-plasma, vapor-2}.

Combining the $T_s$-dependence with the kinetic variations due to the change of $I$ and $\rho$ (see Figs. 4 and 5), we conclude that the optical swelling of deposited nanocarbons does not necessarily require the existence of microbubbles, and that a key candidate of swelling factors is optofluidic migration, such as thermophoresis of nanocarbons, induced by the thermal gradient \cite{thermo-piazza,thermo-gold, thermo-review, thermo-sano, thermo-maeda};
however, an elaborate theory remains to be developed based on the recent advances in the theoretical treatments of thermal dynamics for colloids \cite{thermo-piazza,thermo-gold, thermo-review,thermo-opt,thermo-wurger2,thermo-wurger3}, aiming to validate the phenomenological equation (1), and to elucidate why the relaxation times are close as shown in Fig. 3.
In addition, it is an open problem whether microbubbles, which optical microscopy at $T_s=25\,^\circ\mathrm{C}$ had difficulty in detecting, are indispensable for the nucleation of hollow microcapsule (or the optical deposition of CNTs \cite{opt-capsule, deposition-1, deposition-monitor,deposition-2,graphene-deposition, graphene-device1,graphene-device4, graphene-deposition2}) that emerges from the homogeneous dispersion.

Finally, we note a technical advantage of the OT method.
Figures 1 and 2, as well as Supplementary Movies 1 and 2, have demonstrated that the OT method is able not only to fabricate nanocarbon microcapsule arrays of required configuration, but also to manipulate and deform a single microcapsule.
The use of programmed OT system would be helpful for fabricating more complicated configurations than the triangular arrays of microcapsules (see Fig. 2(b)), though any polymerization in the shells is further required in order to prevent the dried microcapsules from collapsing.
Furthermore, the OT system is capable of accomplishing an alternative LbL assembly:
a CNT microcapsule can be easily covered with a graphene shell by simply exchanging the surrounding suspension.
We envisage the use of the OT method for creating various arrays of hybrid nanocarbon microcapsules that meet engineering demands \cite{hybrid}.

\subsection*{Methods}

{\bf Materials}.
We used the aqueous suspensions of carboxylated multi-walled CNTs (3 wt\%, Cluster Instruments), which was diluted to densities of 0.05 wt\% and 0.025 wt\%.
By contrast, to prepare aqueous 0.025 wt\% dispersions of graphite, we wrapped the carbon dispersoids with DNA via the following procedure, the same steps in preparing aqueous solution from unmodified CNT soot:
The graphite powder (2 $\mu$m in average size, Ito Graphite) was suspended by sonication in an aqueous solution with 0.025-wt\% salmon DNA (Wako) that was denatured by heating for ten minutes at $90\,^\circ\mathrm{C}$.
Subsequently, the suspensions were centrifuged at 650g for ten minutes to extract the carefully decanted supernatants.
We also used a graphene suspension (Nanointegris) with a density of 5 wt\%.
For all of the experiments, a 100-$\mu$l drop of the suspension was mounted on an ordinary cover slip (NEO, Matsunami).
No seals were used for eliminating a boundary effect on the swelling processes, so that microcapsules could be manipulated using microneedles;
incidentally, we verified that the open densification of the drops due to optical heating for five minutes yielded a maximal change in the density of the suspension of less than 10 \%.
The spherical colloids used are carboxylate-modified polystyrene particles (3 $\mu$m in diameter, Dainichi Seika), which were included with the nanocarbon microcapsules to track Brownian motions of the confined particles.

{\bf Experimental setup}.
The OT system (MMS, Sigma Koki) is composed of an inverted optical microscope (TE2000-U, Nikon) with an objective lens (Plan Fluor ELWD $40\times/0.60$, Nikon), a cw Nd:YAG laser ($\lambda=1064\,\mathrm{nm}$), and other optical devices including a CCD camera (Retiga Exi, Q-Imaging) and a fast CMOS camera (PL-A741, Pixelink).
The microscope stage was heated with a silicone rubber electric heating sheet and was maintained at  either $25\,^\circ\mathrm{C}$ or $60\,^\circ\mathrm{C}$ using a temperature controller (DB1000, CHINO).
The light intensity at the focal point ranged from 5 mW to 300 mW, and the longest laser emission duration was 250 seconds.
To manipulate the hollow microcapsules, we used a tungsten microneedle with a tip diameter of 500 nm under the control of a patch clamp micromanipulator (NMN-21, Narishige).


\begin{table*}[htbp]
\caption{({\bf Supplementary Table S1}): Fitting results for CNT suspensions with different densities $\rho$ at laser powers $I$ of 47 mW and 140 mW.
The symbols $t_0/\tau$, $\tau$ and $D_{\mathrm{max}}$ designate the same quantities as those in Table S1.
}
{\renewcommand{\arraystretch}{1.6}
\begin{tabular}{l||cc|cc}
\hline
\phantom{a}$I$ [mW], $\rho$ [wt\%]\phantom{a}&\phantom{a}$t_0/\tau$[s]\phantom{a}&\phantom{a}$\tau$ [s]\phantom{a}&\phantom{a}$D_{\mathrm{max}}/\tau$ [$\mu$m/s]\phantom{a}&\phantom{a}$D_{\mathrm{max}}$ [$\times10^{-4}$ m]\\
\hline
\phantom{ab}47,\phantom{a}0.025&0.2&15&16&2.4\\
\phantom{ab}47,\phantom{a}0.05&0.2&18&18&3.3\\
\phantom{ab}47,\phantom{a}0.1&0.2&20&17&3.3\\
\phantom{ab}140,\phantom{a}0.025&0.3&17&26&4.4\\
\phantom{ab}140,\phantom{a}0.05&0.2&15&34&5.1\\
\hline
\end{tabular}}
\end{table*}

\begin{table*}[htbp]
\caption{({\bf Supplementary Table S2}): Fitting results for graphene and graphite suspensions with different densities at laser powers $I$ of 47 mW, 140 mW and 240 mW.
The symbols $t_0/\tau$, $\tau$ and $D_{\mathrm{max}}$ denote the same quantities as those in Table S1, whereas $\rho$ represents the density of the graphite or graphene solution instead of the CNT solution.}
{\renewcommand{\arraystretch}{1.6}
\begin{tabular}{l|c||cc|cc}
\hline
\phantom{a}$I$ [mW], $\rho$ [wt\%]\phantom{a}&\phantom{a}solutes&\phantom{aa}$t_0/\tau$[s]\phantom{a}&\phantom{a}$\tau$ [s]\phantom{a}&\phantom{a}$D_{\mathrm{max}}/\tau$ [$\mu$m/s]\phantom{a}&\phantom{a}$D_{\mathrm{max}}$ [$\times10^{-4}$ m]\\
\hline
\phantom{ab}140,\phantom{a}5&\phantom{aa}graphene\phantom{aa}&0.2&20&12&2.4\\
\phantom{ab}47,\phantom{a}0.05&graphite&0.3&21&7.0&1.6\\
\phantom{ab}140,\phantom{a}0.05&graphite&0.3&20&16&3.2\\
\phantom{ab}240,\phantom{a}0.05&graphite&0.3&21&20&4.9\\
\phantom{ab}240,\phantom{a}0.025&graphite&0.2&18&20&3.6\\
\hline
\end{tabular}}
\end{table*}

\begin{table*}[htbp]
\caption{({\bf Supplementary Table S3}): Fitting results of the normalized emergence time $t_0/\tau$, the relaxation time $\tau$, and the maximum diameter $D_{\mathrm{max}}$ at saturation for CNT suspensions ($\rho=0.05$ wt$\%$) at different solvent temperatures of $T_s=25\,^\circ\mathrm{C}$ and $T_s=60\,^\circ\mathrm{C}$ while maintaining the incident laser power $I$ at $I=96$ mW.
The symbol $\eta$ in this table represents the water viscosity.}
{\renewcommand{\arraystretch}{1.6}
\begin{tabular}{c|c||cc|ccc}
\hline
$T_s [^\circ\mathrm{C}]$&\phantom{ab}$\eta$ [mPa$\cdot$s]\phantom{a}&\phantom{a}$t_0/\tau$[s]\phantom{a}&\phantom{a}$\tau$ [s]\phantom{a}&\phantom{a}$D_{\mathrm{max}}/\tau$ [$\mu$m/s]\phantom{a}&\phantom{a}$D_{\mathrm{max}}$ [$\times10^{-4}$ m]\\
\hline
\phantom{a}25\phantom{a}&\phantom{a}0.894\phantom{a}&0.2&17&21&3.7\\
\phantom{a}60\phantom{a}&\phantom{a}0.467\phantom{a}&0.3&8&24&2.0\\
\hline
\end{tabular}}
\end{table*}

\end{document}